\documentclass[runningheads, a4paper]{llncs}
\usepackage{graphicx}
\usepackage[linesnumbered,lined,boxed,commentsnumbered]{algorithm2e}
\usepackage{amssymb}
\usepackage{amsmath}
\usepackage{mathrsfs}
\usepackage{mathtools}
\usepackage[pdfpagelabels]{hyperref}
\usepackage{pifont}
\usepackage{fontawesome}
\usepackage{stmaryrd}
\usepackage{enumerate}
\usepackage{multirow}
\usepackage{etoolbox}
\usepackage{graphicx}
\usepackage{gensymb}
\usepackage{centernot}
\usepackage[dvipsnames]{xcolor}
\usepackage{tikz}
\usepackage{tkz-euclide}
\usetikzlibrary{decorations.shapes}
\usepackage{appendix}
\usetikzlibrary{arrows,shapes.geometric}
\usetikzlibrary{calc}
\usepackage{caption}
\usepackage{subcaption}

\hypersetup{
	colorlinks=true,
	linkcolor=blue,     
	urlcolor=cyan}

\usepackage[strings]{underscore}

 \DeclareMathOperator*{\Int}{int}

 \DeclareMathOperator*{\Cl}{cl}

\begin{document}
\title{MITL Verification Under Timing Uncertainty\thanks{The research for this paper received funding from the Australian Government through Trusted Autonomous Systems, a Defence Cooperative Research Centre funded through the Next Generation Technologies Fund.}}
%
%
\author{Daniel Selvaratnam\inst{1} \and
Michael Cantoni\inst{1} \and
J. M. Davoren\inst{1} \and
Iman Shames\inst{2}}
\authorrunning{D. Selvaratnam et al.}
%
\institute{Department of Electrical and Electronic Engineering, The University of Melbourne, Parkville, VIC 3010, Australia \and
CIICADA Lab, School of Engineering, Australian National University, Canberra, ACT 2600, Australia}
\maketitle              
\begin{abstract}
A Metric Interval Temporal Logic (MITL) verification algorithm is presented. It verifies continuous-time signals without relying on high frequency sampling. Instead, it is assumed that collections of over- and under-approximating intervals are available for the times at which the individual atomic propositions hold true for a given signal. These are combined inductively to generate corresponding over- and under-approximations for the specified MITL formula. The gap between the over- and under-approximations reflects timing uncertainty with respect to the signal being verified, thereby providing a quantitative measure of the conservativeness of the algorithm. The verification is exact when the over-approximations for the atomic propositions coincide with the under-approximations. Numerical examples are provided to illustrate.

\end{abstract}
%
%
%

\newcommand\R{\mathbb{R}}
\newcommand\C{\mathbb{C}}
\newcommand\Q{\mathcal{Q}}
\newcommand\N{\mathbb{N}}
\newcommand\II{\mathbb{I}}
\newcommand\IQ{\II_\mathbb{Q}}
\newcommand\QQ{\mathfrak{Q}}

\renewcommand\P{\mathcal{P}}
\newcommand\K{\mathcal{K}}
\newcommand\union{{\textstyle \bigcup}}
\newcommand\F{\mathcal{F}}
\newcommand\T{\mathcal{T}}
\newcommand\I{\mathcal{I}}
\renewcommand\H{\mathcal{H}}
\newcommand\U{\mathcal{U}}
\newcommand\V{\mathcal{V}}
\newcommand\J{\mathcal{J}}
\newcommand\M{\mathcal{M}}
\renewcommand\O{\mathcal{O}}
\newcommand\B{\mathcal{B}}
\newcommand\A{\mathcal{A}}
\newcommand\W{\mathcal{W}}
\newcommand\D{\mathbf{D}}
\newcommand\x{\mathbf{x}}
\newcommand\X{\mathcal{X}}
\newcommand\Y{\mathcal{Y}}
\renewcommand\v{\mathbf{v}}
\renewcommand\u{\mathbf{u}}
\newcommand\y{\mathbf{y}}
\newcommand\z{\mathbf{z}}
\newcommand\w{\mathbf{w}}
\newcommand\0{\mathbf{0}}
\newcommand\1{\mathbf{1}}

\newcommand\tL{\mathtt{Left}}
\newcommand\tR{\mathtt{Right}}
\newcommand\tS{\mathtt{Seek}}
\newcommand\tO{\mathtt{Stop}}
\newcommand\g{\mathtt{g}}
\newcommand\G{\mathtt{G}}

\newcommand\diag{\mathrm{diag}}
\newcommand\spn{\mathrm{span}}
\newcommand\st{\text{ s.t. }}
\newcommand\IF{\text{if }}
\newcommand\dom{\text{dom }}
\newcommand\dashV{\mathrel{\reflectbox{\ensuremath{\vDash}}}}

\newcommand\until{\mathscr{U}}
\newcommand\always{\square}
\newcommand\eventually{\lozenge}
\newcommand\release{\mathsf{R}}
\newcommand\weak{\mathsf{W}}
\renewcommand\next{\bigcirc}
\newcommand\limplies{\rightarrow}
\newcommand\liff{\leftrightarrow}
\newcommand\oleft{\text{\faArrowLeft}}
\newcommand\oright{\text{\faArrowRight}}
\newcommand\obstacle{\text{\ding{54}}}
\newcommand\otarget{\text{\ding{52}}}

\newcommand\proj{ \mathsf{\Pi}}
\newcommand\cont{ \mathsf{\Lambda}}
\newcommand\Prop{ \mathrm{Prop}}
\newcommand\LTL{ \mathrm{LTL}}
\newcommand\MITL{ \mathfrak{F}}
\newcommand\symdiff{\mathbin{\triangle}}

\section{Introduction}
Metric Interval Temporal Logic (MITL), introduced in \cite{alurBenefitsRelaxingPunctuality1996}, is a logic for specifying behaviours of continuous-time signals. Given such a signal, the problem of interest is to test whether it satisfies a given MITL formula. Also known as \emph{runtime verification} or \emph{monitoring}, this has been called the \emph{path checking} problem in \cite{basinOptimalProofsLinear2018,selvaratnamSamplingPolynomialTrajectories2022}.
Since the domain of a continuous-time signal is uncountably infinite, the design of a general MITL path checking algorithm is challenging. In practice, the times at which even well-behaved signals change value cannot be known with infinite precision. In various situations, it may still be possible to obtain upper and lower bounds on the times at which the individual atomic propositions switch between true and false. The verification of MITL formulas, given such bounds, is the topic of this paper. The proposed algorithm is intended for offline verification. 

An algorithm for path checking continuous-time signals against MITL specifications is proposed in \cite{malerMonitoringTemporalProperties2004}. The algorithm operates directly on the time intervals over which the individual propositions are true. Exact knowledge of these intervals would suffice for a correct verification result. With such knowledge being unrealistic, the authors assume a sampling strategy, for the purpose of implementation, that is dense enough to capture every change over time to the set of atomic propositions true at each instant. Such a lossless sampling strategy remains impossible to guarantee in general. Our approach is to start instead with under- and over-approximations of those intervals, and combine them appropriately for the verification of compound formulas. For the purpose of analysis, we utilise the topological concepts of separation and connectedness. This formal approach enables us to present rigorous proofs, and to extend \cite{malerMonitoringTemporalProperties2004} in other directions as well. Specifically, \cite{malerMonitoringTemporalProperties2004} restricts all the atomic proposition intervals to be left-closed right-open, and all the operator timing bounds to be closed and bounded. We remove all of these restrictions, but for finiteness, still require the truth sets of the atomic propositions to be expressible as the union of a finite number of (possibly unbounded) intervals. Moreover, our procedure accommodates the \emph{strict non-matching} until operator, which is more expressive~\cite{furiaExpressivenessMTLVariants2007} than the \emph{non-strict matching} until adopted in \cite{malerMonitoringTemporalProperties2004}. For this reason, the strict non-matching until is taken as primitive in the standard semantics of MITL~\cite{alurBenefitsRelaxingPunctuality1996}. The developments herein make no modifications to those original semantics.

Other notable attempts to relax the lossless sampling assumption of \cite{malerMonitoringTemporalProperties2004} include \cite{fainekosRobustnessTemporalLogic2009} and \cite{furiaTheorySamplingContinuoustime2010}. The former relies on high-frequency sampling and spatial robustness margins in order to infer continuous-time satisfaction from discrete-time satisfaction. In addition, it requires a tightening of the sub-formulas at every level of the parse tree. The latter also relies on high-frequency sampling, and restricts attention to flat formulas, which do not have nested temporal operators. Interestingly, \cite{furiaTheorySamplingContinuoustime2010} also constructs over- and under-approximations to the truth, but in discrete-time. 
Our approach differs from \cite{fainekosRobustnessTemporalLogic2009,furiaTheorySamplingContinuoustime2010} by not relying on high-frequency sampling. We also place no restrictions on, and make no modifications to, the original formula to be verified.

Instead of relaxing the lossless sampling assumption of \cite{malerMonitoringTemporalProperties2004}, the approach in \cite{selvaratnamSamplingPolynomialTrajectories2022} ensures it by restricting the scope to continuous-time polynomial splines. Root isolation techniques from computer algebra are invoked to ensure that every change to the set of true atomic propositions is recorded. The formulas there are restricted to LTL without next, which is introduced as a fragment of MITL with only unbounded temporal operators. This paper can be viewed as a sequel to \cite{selvaratnamSamplingPolynomialTrajectories2022} that considers bounded temporal operators. The restricted setting in \cite{selvaratnamSamplingPolynomialTrajectories2022} yields a verification algorithm that always delivers a conclusive result. Due to the timing uncertainty accommodated here, some conservativeness is unavoidable: the outcome may be inconclusive. In view of this, our path checking algorithm quantifies how much the timing uncertainty in the atomic propositions may be amplified as it propagates through the parse tree of the formula.
If the truth intervals for the atomic propositions are known exactly, then the algorithm also constructs the intervals for the formula exactly. More generally, the timing uncertainty associated with satisfying the formula relates to the gap between the over- and under-approximations produced by the algorithm. Although its precise nature is deferred to future work, we anticipate a relationship between this gap and the notions of temporal robustness in \cite{donzeRobustSatisfactionTemporal2010,lindemannTemporalRobustnessStochastic2022,rodionovaTemporalRobustnessTemporal2022}. Such is of relevance to recent works~\cite{linOptimizationbasedMotionPlanning2020,rodionovaTimeRobustControlSTL2021}, which aim to synthesize controllers that maximise temporal robustness. 

Our approach relies on starting with over- and under-approximations to the truth sets of the atomic propositions. The following two scenarios provide examples of how such approximations may be generated. First, consider the verification of a continuous-time signal that passes through different state-space regions, each corresponding to an atomic proposition. 
In the polynomial setting of \cite{selvaratnamSamplingPolynomialTrajectories2022}, root isolation algorithms provide samples on either side of every region transition, even though the transition points cannot be found exactly. This information translates directly to the over- and under-approximating time intervals assumed here. Second, suppose that the regions themselves are not known precisely, but outer and inner boundaries for them are known instead. Time spent within the outer boundaries correspond to over-approximations, and the inner-boundaries to under-approximations. 
\section{Preliminaries} \label{sec:prelim}
\subsection{Set theory} \label{sec:sets}
The subset relation is denoted by $\subseteq$, and the strict subset relation by $\subsetneq$. Let $|A|$ be the cardinality of set $A$, and $2^A$ its power set. If $f:A \to B$, then let $f[X] \subseteq B$ denote the image of $X \subseteq A$, and $f^{-1}Y \subseteq A$ the preimage of $Y \subseteq B$.
The operator $\union$, when used without limits, takes the union of all the elements within a set of sets. It is then given binding precedence over all other set operations. Thus, $\union \P := \bigcup_{B \in \P} B,$ and $\union \P \cap \union \Q = (\union \P) \cap (\union \Q)$, where the elements of $\P$ and $\Q$ are sets. For sequences of sets of sets, note that $ \union \I_k = \bigcup_{I \in \I_k} I \neq \bigcup_k \I_k.$ Let $\R$ denote the real numbers, $\mathbb{Q}$ the rational numbers, $\mathbb{Z}$ the integers, and $\mathbb{N} := \{0,1,...\}$ the natural numbers. For $i,j \in \mathbb{Z}$, let $ [i:j]:= \{k \in \mathbb{Z} \mid i \leq k \leq j \}$. Analogous definitions apply for $[i:j), (i:j]$ and $(i:j)$.  
\subsection{Topology} \label{sec:topology}
We rely on the following facts from topology. Given a subset $A$ of a topological space, let $\Int(A)$ denote its interior, and $\Cl(A)$ its closure.  Recall that any subset of a topological space is itself a topological space when endowed with the subspace topology.
\begin{proposition} \label{prop:interiorSub}
	Let $A,B$ be subsets of a topological space. If $A \subseteq B$, then $\Cl(A) \subseteq \Cl(B)$ and $\Int(A) \subseteq \Int(B)$. 
\end{proposition}
A topological space is \emph{connected} iff it is not the union of two of its disjoint open subsets \cite[\S 23]{munkresTopology2000}. Two subsets $A,B$ are \emph{separated} iff $A \cap \Cl(B) = B \cap \Cl(A) = \emptyset$. Clearly, if $A$ and $B$ are separated, then they are disjoint, and furthermore, any $C \subseteq A$ and $D \subseteq B$ are also separated. The result below is from \cite[Theorem 6.1.1]{engelkingGeneralTopology1989}.
\begin{proposition} \label{prop:union of separated subsets}
		Let $A$ and $B$ be non-empty subsets of a topological space. If $A$ and $B$ are separated, then $A \cup B$ is not connected. 
\end{proposition}
\begin{proposition} \label{prop:connected}
	Let $A$ and $B$ be separated subsets of a topological space. Suppose $ \emptyset \neq S \subseteq A \cup B$. If $S$ is connected, then either $S \subseteq A$ or $S \subseteq B$, but not both.
\end{proposition}
\begin{proof}
	That $S \subseteq A$ or $S \subseteq B$ is established in \cite[Corollary 6.1.8]{engelkingGeneralTopology1989}. If $S \subseteq A \cap B$, then $A$ and $B$ intersect, which contradicts the assumption that they are separated. 
\qed \end{proof}
\begin{proposition} \label{prop:separation from union}
	Let $A, B_1,...,B_m$ be subsets of a topological space. If for every $i \in [1:m]$, $A$ and $B_i$ are separated, then $A$ and $\bigcup_{i=1}^m B_i$ are separated. 
\end{proposition}
\begin{proof}
	First observe that $\Cl(A) \cap B_i = \emptyset$ for all $i \in [1:m]$, 
	which implies
	$ \Cl(A) \cap \bigcup_{i=1}^m B_i = \emptyset.$
	It is also true that $A \cap \Cl(B_i) = \emptyset$ for all $i \in [1:m]$, which implies
	$ A \cap \Cl \left( \bigcup_{i=1}^m B_i \right) = A \cap \bigcup_{i=1}^m \Cl(B_i) = \emptyset,$
	because the finite union of closures is the closure of their union \cite[Theorem 1.1.3]{engelkingGeneralTopology1989}.
\qed \end{proof}
In what follows, we work mostly with subsets of the real line $\R$, which is a connected topological space. Of particular interest are subsets of the non-negative reals. However, it must be noted that for any $A \subseteq [0,\infty)$, $\Int(A)$ denotes the interior of $A$ relative to $\R$, not $[0,\infty)$. 
\begin{definition}
	An \emph{interval} is a connected subset of the real line.  
\end{definition} 
An equivalent definition is that intervals are the convex subsets of the real line. An interval $I$ is \emph{degenerate} iff $|I| \leq 1$. Otherwise, it is \emph{nondegenerate}. 
Let $\II := \{ I \subseteq [0,\infty) \mid I \text{ is an interval}\},$ which is the set of non-negative real intervals, and
$$ \II_\mathbb{Q} := \{ I \in \II \setminus \{ \emptyset \} \mid \sup I, \inf I \in \mathbb{Q} \cup \{ \infty \} \}, $$
the set of non-empty non-negative real intervals with rational endpoints.
\begin{proposition} \label{prop:interiorOfClosure}
For any interval $I$, $\Int(\Cl(I)) \subseteq I$.
\end{proposition}
\begin{proof} 
	Taking the closure of an interval adds to it at most two points, which are its endpoints. The result remains an interval with the same endpoints. Taking the interior of that then simply removes the endpoints. No point is added by the closure that is not removed by the interior. 
	\qed \end{proof}
\section{Problem formulation} \label{sec:prob}
\subsection{Syntax and semantics of MITL}
We start with the syntax of MITL~\cite[Definition 2.3.1.1]{alurBenefitsRelaxingPunctuality1996}. Note that while $\II$ is uncountable, $\IQ$ is countable --- a necessary requirement for MITL to be a formal language. 
\begin{definition}[MITL] \label{def:MITL}
	Given a finite set of atomic propositions $\G$,	$\varphi$ is an MITL formula over $\G$ iff it is generated by the grammar
	$$ \varphi::= \top \mid \g \mid \lnot \varphi \mid \varphi \land \varphi \mid \varphi \until_I \varphi,$$
	where $\g \in \G$ and $I \in \IQ$ must be non-degenerate. The set of MITL formulas over $\G$ is $\MITL(\G)$. 
\end{definition} 
The semantics of MITL in \cite[Definition 2.3.2.1]{alurBenefitsRelaxingPunctuality1996} is stated in terms of \emph{timed state sequences} and \emph{interval sequences}. Since we do not make use of such machinery explicitly, we provide a common restatement of the semantics instead, that is equivalent for signals of \emph{finite variability}. Such signals cannot change values infinitely often within a finite time interval. Formally, a function $z:[0,\infty) \to A$ is of finite variability iff there exists an interval sequence~\cite[Definition 2.2.1]{alurBenefitsRelaxingPunctuality1996} $I_1,I_2,\hdots$ such that $z$ is constant over $I_i$ for every $i \in \N$.
\begin{definition} \label{def:MITLsemantics}
	Let $z:[0,\infty) \to 2^\G$ be a finite variability signal, with $\G$ a finite set of atomic propositions. The satisfaction relation $\models$ is defined inductively as follows: Given any $t \geq 0$, proposition $\g \in \G$, formulas $\varphi_1,\varphi_2 \in \MITL(\G)$, and non-degenerate $I \in \IQ$,
	\begin{align} 
		(z,t) \models \top&;  \\
		(z,t) \models \g & \iff \g \in z(t); \label{eq:gDef}  \\
		(z,t) \models \lnot \varphi_1 & \iff (z,t) \not\models \varphi_1; \label{eq:notDef}  \\
		(z,t) \models \varphi_1 \land \varphi_2 & \iff (z,t) \models \varphi_1 \text{ and } (z,t) \models \varphi_2; \label{eq:andDef}  \\
		(z,t) \models \varphi_1 \until_I \varphi_2 &\iff \nonumber \\
		 \exists t_2 \in I,\ &\Big( (z,t + t_2) \models \varphi_2 \text{ and } \forall t_1 \in (0,t_2),\  (z,t+t_1) \models \varphi_1 \Big). \label{eq:untilDef} 
	\end{align}
\end{definition}
The variant of the until operator in Definition \ref{def:MITLsemantics} is called the \emph{strict non-matching} until. There are three other variants, corresponding to the other possible permutations of including or excluding the endpoints of $(0,t_2)$ in \eqref{eq:untilDef}. All other variants can be expressed simply in terms of the strict non-matching until (see \cite{furiaExpressivenessMTLVariants2007}), and it is therefore taken as primitive.
It is now possible to formally state the problem of interest.
\begin{problem} \label{prob:basic}
	Given a finite set of atomic propositions $\G$, formula $\varphi \in \MITL(\G)$, finite variability signal $z:[0,\infty) \to 2^\G$, and time $t \in [0,\infty)$, determine whether ${(z,t) \models \varphi}$. 
\end{problem}
\subsection{Truth sets and interval queues}
We are interested in tackling Problem \ref{prob:basic} without precise knowledge of the times at which $z$ changes value. To describe this scenario mathematically, we introduce the truth set of a formula.
\begin{definition}[Truth set]
	Given a finite set of atomic propositions $\G$, and a finite-variability signal $z:[0,\infty) \to 2^\G$, the \emph{truth set} of $\varphi \in \MITL(\G)$ is given by
	$$T_z(\varphi):= \{ t \in [0,\infty) \mid (z,t) \models \varphi \}.$$ 
\end{definition} 
By definition, $(z,t) \models \varphi$ if and only if $t \in T_z(\varphi)$. The obvious strategy for solving Problem \ref{prob:basic} is to construct the truth set of a formula from the truth sets of the atomic propositions that appear in the formula. In general, truth sets may be quite complex, and they need not have a finite representation because $z$ has an infinite duration. Let us initially restrict attention to problems where the truth sets of the atomic propositions can be written as the union of a finite set of non-empty separated intervals, which we call an interval queue. (Remark \ref{rem:infiniteSwitching} later explains how this restriction can be relaxed.)
\begin{definition}[Interval queue] \label{def:interval Q}
	An \emph{interval queue} is a finite set $\I \subseteq \II \setminus \{ \emptyset \}$, such that for any $I,J \in \I$,
	$ I \neq J \implies I \text{ and } J \text{ are separated.} $
Let $\QQ$ denote the set of interval queues.
\end{definition}
Thus, an interval queue is a finite set of non-empty non-negative separated real intervals. Note that separation here allows for `isolated holes': $\{ [0,1),(1,2]\}$ is an interval queue, for example. 
Although any interval queue is a finite set by definition, its elements may still be unbounded intervals. Also, note that $\emptyset$ is an interval queue, but $\{ \emptyset \}$ is not. 
If we have interval queue representations for the truth sets of the atomic propositions, then Problem \ref{prob:basic} reduces to the one below. 
\begin{problem} \label{prob:exact}
	Let $\G$ be a finite set of atomic propositions, $z:[0,\infty) \to 2^\G$ a finite variability signal, and $\varphi \in \MITL(\G)$. Given $\P:\G \to \QQ$ such that $ \forall \g \in \G,\ \union \P(\g) = T_z(\g),$
	construct an interval queue $\Q \in \QQ$ such that $\union \Q = T_z(\varphi)$. 
\end{problem}
Above, $\P$ assigns to each atomic proposition an interval queue representation of its truth set. To accommodate our lack of precise timing knowledge, we now introduce maps $\P^-$ and $\P^+$, that provide under- and over-approximating interval queues, respectively, for the truth set of every atomic proposition.
\begin{problem} \label{prob:relaxed}
	Let $\G$ be a finite set of atomic propositions, $z:[0,\infty) \to 2^\G$ a finite variability signal, and $\varphi \in \MITL(\G)$. Given $\P^-:\G \to \QQ$ and $\P^+:\G \to \QQ$ such that
		\begin{equation} \forall \g \in\G,\ \union \P^-(\g) \subseteq T_z(\g) \subseteq \union \P^+(\g), \label{eq:relaxedPs}
		 \end{equation}
	 construct an under-approximating $\Q^- \in \QQ$ and over-approximating $\Q^+ \in \QQ$ such that $$\union \Q^- \subseteq T_z(\varphi) \subseteq \union \Q^+.$$
\end{problem}
An algorithm that solves Problem \ref{prob:relaxed} is developed in Section \ref{sec:solution} using foundations established in Section \ref{sec:basics}. In the special case that $\P^- = \P^+$ (i.e., the under- and over-approximations to the truth sets of the atomic propositions are equal), the algorithm solves Problem \ref{prob:exact} exactly. 
\begin{remark}[Finite information horizons] \label{rem:finiteHor}
In practice, information about the signal $z:[0,\infty) \to 2^\G$ may not be available infinitely far into the future. Suppose $z(t)$ is unknown for all $t > b$, where $b \geq 0$ represents the horizon of available information. To accommodate this, for every atomic proposition, the interval $ (b,\infty)$ must be contained in the over-approximation of its truth set, but not in the under-approximation. That is,
$$ (b,\infty) \subseteq \bigcap_{\g \in \G} \left( \union \P^+(\g) \setminus \union \P^-(\g) \right). $$
\end{remark}
\begin{remark} \label{rem:infiniteSwitching}
Even for a finite variability signal $z:[0,\infty) \to 2^\G$, the truth set $T_z(\g)$ of atomic proposition $\g \in \G$ may not have an interval queue representation, because $\g$ may switch between true and false infinitely often over infinite time. Even so, it is always possible to construct interval queue under- and over-approximations for its truth set, by artificially imposing a finite information horizon $b \geq 0$, as described in Remark \ref{rem:finiteHor}, but this time only requring $ (b,\infty) \subseteq \union \P^+(\g) \setminus \union \P^-(\g)$ to hold for the individual proposition $\g$. For example, if $T_z(\g) = \bigcup_{k=0}^\infty [k, k + \frac{1}{2}]$, choose $b \in \N$ positive, and set $\P^-(\g) = \{ [k, k + \frac{1}{2}] \mid k < b \}$ and $\P^+(\g):= \P^-(\g) \cup \{ [b,\infty) \}$. This effectively `truncates' the truth set. 
	\end{remark}
\begin{remark}[Endpoint rationality] If an interval queue $\Q \subseteq \IQ$, then all of its elements have rational endpoints, and it admits a finite representation. The mathematical procedures that follow do not rely on endpoint rationality for their correctness, but it is of course vital for their algorithmic implementation. For simplicity then, results are stated for general interval queues, while noting alongside that all the necessary operations preserve endpoint rationality. 
	\end{remark}
\begin{remark}[Uniqueness]
	If an interval queue representation of a set exists, then it is unique. This fact follows from the separation of the individual intervals, but since the following results do not exploit uniqueness, it is not proved here. 
\end{remark}

\section{Interval queue fundamentals} \label{sec:basics}
Our solution to Problem \ref{prob:relaxed} features two key subroutines, which are discussed here. The first constructs an interval queue from a given finite set of (possibly overlapping) non-negative intervals, and the second an interval queue that complements another interval queue. 
\subsection{Construction algorithm}
Algorithm \ref{alg:interval queue} simply merges any intervals that are adjacent or overlap. The set $\J$ maintains a collection of separated intervals. Every starting interval $I_i$ is either added to $\J$ without modification, or merged with existing intervals in $\J$. Given a new candidate interval $I_i$, the set $\W$ in Line \ref{ln:Wi} selects both it and all intervals in $\J$ that are connected to it. In Line \ref{ln:Ji}, these intervals are removed from $\J$, and replaced with their union, which is necessarily an interval.

\begin{algorithm}[h]
	\SetAlgoLined
	\DontPrintSemicolon
	\SetKwFunction{FnFormIQ}{ConstructIQ}
	\SetKwProg{Fn}{Function}{:}{end}
	\KwIn{$I_1,...,I_m \in \II$}
	\KwOut{$\J \in \QQ$ such that $\union \J = \bigcup_{i=1}^m I_i$}
	\BlankLine
	\Fn{\FnFormIQ{$\{I_1,...,I_m\}$}}{
		$\J := \emptyset$\;
		\For{ $i \in [1:m]$}{$\mathcal{W} := \{J \in \J \mid I_i \cup J \in \II \} \cup \{I_i\}$ \label{ln:Wi} \;
			$\J \leftarrow (\J \setminus \mathcal{W}) \cup \left\{ \union \mathcal{W} \right\}$ \label{ln:Ji}\;
		}
		\KwRet $\J \setminus \{ \emptyset \}$}
	\caption{constructs an interval queue from a given finite set of non-negative intervals.}
	\label{alg:interval queue}
\end{algorithm}

\begin{lemma} \label{lem:FormIQ}
	For any finite set $\I \subseteq \II$, $\mathtt{ConstructIQ}(\I)$ is an interval queue such that $\union \mathtt{ConstructIQ}(\I) = \union \I$. Moreover, $\I \subseteq \IQ \implies \mathtt{ConstructIQ}(\I) \subseteq \IQ$. 
\end{lemma}

\subsection{Complementation algorithm}
Given an interval queue $\I \in \QQ$, we now seek another interval queue $\J \in \QQ$ that complements it: $\union \J = [0,\infty) \setminus \union \I$. If $\I:=\{ (1,2],(3,4] \}$, then clearly its complement is
$ \J:=\{ [0,1], (2,3], (4,\infty) \} $. Although the intervals in $\I$ are all bounded left-open right-closed, the complementary queue $\J$ contains, in addition, intervals that are closed, open, and unbounded. Given the possible types of end-points, there are 6 different types of intervals that may arise. Considering each case in turn, and their pairwise combinations, threatens to be an excessively laborious task. To avoid this, we use notation that accommodates all of them within a unified framework, and allows our complementation algorithm to be stated succinctly. 
Given $A,B \subseteq [0,\infty)$, define
\begin{align}	
	\overrightarrow{B} &:=  \{ t \geq 0 \mid \forall b \in B,\ t > b \} = \bigcap_{b \in B} (b,\infty) \\
	\overleftarrow{A} &: = \{ t \geq 0 \mid \forall a \in A,\ t < a  \} = \bigcap_{a \in A} [0,a).
\end{align}
By the above definitions, $\overleftarrow{\emptyset} = \overrightarrow{\emptyset} = [0,\infty)$. 
If $A,B \neq \emptyset$, then $\overrightarrow{A}$ is an interval containing everything strictly to the right of $A$, and $\overleftarrow{B}$ is an interval containing everything strictly to the left of $B$. 
For example, given $0 < a < b,$ $\overrightarrow{(a,b)} = [b,\infty)$, $\overleftarrow{(a,b)} = [0,a]$, $\overrightarrow{(a,b]} = (b,\infty)$, $\overleftarrow{[a,b)} = [0,a)$, $\overrightarrow{(a,\infty)} = \emptyset$, and $\overleftarrow{[0,b]} = \emptyset$.

Define the binary `earlier-than' relation $\prec$ between non-empty subsets of $[0,\infty)$ as follows: $A \prec B$ iff 
$ \forall a \in A,\ \forall b \in B,\ a < b.$
That is, $A \prec B \iff  B \subseteq \overrightarrow{A}$, for $A,B \neq \emptyset$.  
This relation is a strict partial order on $2^{[0,\infty)}$.

 The complementation algorithm is presented in Algorithm \ref{alg:complementary interval queue}. It yields separated outputs because the elements of the input interval queue lie strictly in-between them. Observe also that $\overrightarrow{(\cdot)}$ and $\overleftarrow{(\cdot)}$ preserve the endpoint rationality of intervals. 
 
\begin{algorithm}[h]
	\SetAlgoLined
	\DontPrintSemicolon
	\SetKwFunction{FnCompIQ}{ComplementIQ}
	\SetKwProg{Fn}{Function}{:}{end}
	\KwIn{$\I \in \QQ$}
	\KwOut{$\J \in \QQ$ such that $\union \J = [0,\infty) \setminus \union \I$}
	\BlankLine
	\Fn{\FnCompIQ{$\I$}}{
		\eIf{$\I = \emptyset $}{ \KwRet $\{ [0,\infty) \}$}{
			Sort $ \I = \{I_1,...,I_m\}$ such that $I_1 \prec I_2 \hdots \prec I_m$ \label{ln:sort} \;
			$\J := \{\overleftarrow{I}_1, \overrightarrow{I}_m \}$ \;
			\For{ $k \in [1:m)$}{
				$\J \leftarrow \J \cup \left\{ \overrightarrow{I}_k \cap \overleftarrow{I}_{k+1} \right\}$}
			\KwRet $\J \setminus \{ \emptyset \}$}} \label{ln:output of ComplementIQ}
	\caption{constructs the complementary interval queue for a given interval queue.}
	\label{alg:complementary interval queue}
\end{algorithm}

\begin{lemma} \label{lem:complement IQ}
	For any interval queue $\I \in \QQ$, $\mathtt{ComplementIQ}(\I) \in \QQ$ is an interval queue such that $\union \mathtt{ComplementIQ}(\I)= [0,\infty) \setminus \union \I$. Moreover, $\I \subseteq \IQ \implies \mathtt{ComplementIQ}(\I) \subseteq \IQ$.
\end{lemma}
\section{Verification procedure} \label{sec:solution}
The goal of this section is to construct maps $\Q^-:\MITL(\G) \to \QQ$ and $\Q^+:\MITL(\G) \to \QQ$, such that, for any formula $\varphi$, $\Q^-(\varphi)$ is the under-approximation and $\Q^+(\varphi)$ the over-approximation that together solve Problem \ref{prob:relaxed}. Then if $t \in \union \Q^-(\varphi)$, we can conclude that $(z,t) \models \varphi$, and $t \notin \union \Q^+(\varphi)$ allows us to conclude that $(z,t) \not\models \varphi$. 
\subsection{Operators in the interval queue domain}
The MITL grammar in Definition \ref{def:MITL} has three primitive operators: complementation; conjunction; and strict non-matching until. The semantics in Definition~\ref{def:MITLsemantics} can be re-expressed in terms of truth sets as follows:
\begin{align}
T_z(\lnot \varphi_1) &= [0,\infty) \setminus T_z(\varphi_1); \label{eq:truthSetNot} \\
T_z(\varphi_1 \land \varphi_2) &= T_z(\varphi_1) \cap T_z(\varphi_2); \label{eq:truthSetAnd} \\
t \in T_z(\varphi_1 \until_I \varphi_2) &\iff \exists t_2 \in I,\ \Big( t + t_2 \in T_z(\varphi_2) \text{ and } (t,t+t_2) \subseteq T_z(\varphi_1) \Big). \label{eq:truthSetUntil}
\end{align}
For each of these logical operators, corresponding operators in the interval queue domain can be defined to generate interval queues that preserve the above truth set relationships. First recall the Minkowski difference between sets $A,B \subseteq \R$ is as follows: $B \ominus A := \{b - a \mid b \in B,\ a \in A\}$.~\footnote{\label{ft:noneg}Cf.  \cite{malerMonitoringTemporalProperties2004}: intersection with $[0,\infty)$ is redundant here, because of the intersection with $\Cl(H) \subseteq [0,\infty)$ in \eqref{eq:U}.}
\begin{definition} \label{def:interval queue operations} The following operations are defined for any $\H, \I, \J \in \QQ$:
	\begin{enumerate}[\thedefinition .1)]
		\item ${\sim} \I:= \mathtt{ComplementIQ}(\I)$;
		\item $\I \sqcap \J :=\{ I \cap J \mid I \in \I,\ J \in \J \} \setminus \{ \emptyset \}$; \label{cl:I and J}
		\item \label{cl:IuntilJ} Given any $I \in \II$, $\H \boxdot_I \J:=  \mathtt{ConstructIQ}(\Y) $, where the set \begin{equation} \Y:= \left\{ \Big(( \Cl(H) \cap J) \ominus I \Big) \cap \Cl(H) \mid H \in \H, J \in \J \right\} . \label{eq:U} \end{equation}
	\end{enumerate}
\end{definition}
\begin{remark}
	The simplicity of the $\sqcap$ operation in Definition \ref{def:interval queue operations} is one reason for choosing conjunction as primitive in Definition \ref{def:MITL}, rather than disjunction.  
\end{remark}
We now show that, if $ \union \I = T_z(\varphi_1)$ and $\union \J = T_z(\varphi_2)$, given $\I,\J \in \QQ$ and $\varphi_1,\varphi_2 \in \MITL(\G)$, then
$T_z(\lnot \varphi_1) = \union ({\sim} \I)$, $T_z(\varphi_1 \land \varphi_2) = \union (\I \sqcap \J)$, $T_z(\varphi_1 \until_I \varphi_2) = \union (\I \boxdot_I \J)$,
and ${\sim} \I, \I \sqcap \J, \I \boxdot_I \J \in \QQ$. Endpoint rationality is also preserved. 
The result for negation is a corollary of Lemma \ref{lem:complement IQ}. Next we turn to conjunction. 

\begin{lemma} \label{lem:intersection IQ}
	If $\I,\J \in \QQ$, then $\I \sqcap \J \in \QQ$ and $\union \left( \I \sqcap \J \right) = \union \I  \cap  \union \J $. Moreover, $\I,\J \subseteq \IQ \implies \I \sqcap \J \subseteq \IQ$.  
\end{lemma}
\begin{proof}
	We first show that $\I \sqcap \J$ is an interval queue. The finiteness of $\I \sqcap \J$ follows from the finiteness of both $\I$ and $\J$. Also, $\emptyset \notin \I \sqcap \J$ by definition. We now establish separation.
	Since $I_1 \cap I_2 \in \II$ for any $I_1,I_2 \in \II$, it follows from Definition \ref{def:interval queue operations} that $\I \sqcap \J \subseteq \II$. 
	Now let $U,V \in \I \sqcap \J$ and suppose that $U \neq V$. Then there exist $I_1,I_2 \in \I$ and $J_1,J_2 \in \J$ such that $U = I_1 \cap J_1$ and $V = I_2 \cap J_2$. Since the closure of a finite intersection is contained in the intersection of the closures \cite[Exercise 1.1.A]{engelkingGeneralTopology1989},
	\begin{align} 
		\Cl(U) \cap V &= \Cl(I_1 \cap J_1) \cap I_2 \cap J_2 \subseteq \Cl(I_1) \cap \Cl(J_1) \cap I_2 \cap J_2, \label{eq:ClUV} \\
		U \cap \Cl(V) &= I_1 \cap J_1 \cap \Cl(I_2 \cap J_2) \subseteq I_1 \cap J_1 \cap \Cl(I_2) \cap \Cl(J_2). \label{eq:UClV}
	\end{align}
	If both $I_1 = I_2$ and $J_1 = J_2$, then $U = V$, which contradicts the hypothesis that $U \neq V$. If $I_1 \neq I_2$, then they are separated because they are elements of an interval queue, and $\Cl(I_1) \cap I_2 = I_1 \cap \Cl(I_2) = \emptyset$, which implies $\Cl(U) \cap V = U \cap \Cl(V) = \emptyset$ in view of \eqref{eq:ClUV} -- \eqref{eq:UClV}. Otherwise $J_1 \neq J_2$, which yields the same result. Thus $U$ and $V$ are separated. As such, $\I \sqcap \J$ is an interval queue.  
	
	Since $\I$, $\J$ are finite, we may assume $\I = \{ I_1,...,I_m\}$ and $\J = \{J_1,...,J_n\}$ without loss of generality. Applying the distributive properties of set unions and intersections,
	$\left( \union \I \right) \cap \left( \union \J \right)  = \left( \bigcup_{i=1}^m I_i \right) \cap \left( \bigcup_{j=1}^n  J_j \right) = \bigcup_{j=1}^n \Big( \left( \bigcup_{i=1}^m I_i \right) \cap J_j \Big) = \bigcup_{j=1}^n \bigcup_{i=1}^m (I_i \cap J_j) = \union( \I \sqcap \J).$
We conclude by noting that the intersection of two intervals with rational endpoints is an interval with rational endpoints. 
\qed 
\end{proof}
The until operation is considerably more complicated than conjunction. The set $\Y$ in \eqref{eq:U}, which is inspired by \cite[Section 3]{malerMonitoringTemporalProperties2004}, may not be an interval queue. However, set intersections, closures, and Minkowski differences preserve convexity (which means that intervals remain intervals) and endpoint rationality. Non-negativity is also preserved, as noted in Footnote \ref{ft:noneg}. Thus, $\Y$ is a finite set of non-negative intervals, and Algorithm \ref{alg:interval queue} is required to ensure that $\H \boxdot_I \J \in \QQ$. Lemma \ref{lem:FormIQ} guarantees that
\begin{equation} \union (\H \boxdot_I \J) = \union \Y = \bigcup_{H \in \H} \bigcup_{J \in \J} \left( \Big(( \Cl(H) \cap J) \ominus I \Big) \cap \Cl(H) \right).  \label{eq:grandUnion} \end{equation}
The lemma below is also of crucial importance to the until operation: if a non-empty interval is contained in the union of an interval queue, then it is fully contained within one of the intervals in that interval queue. 
\begin{lemma} \label{lem:subset of IQ}
	Suppose $S \in \II$ is non-empty, and $S \subseteq \union \I$, with $\I \in \QQ$ an interval queue. Then there exists a unique $I \in \I$ such that $I \supseteq S$. 
\end{lemma}
\begin{proof}
	Let $\I = \{I_1, \hdots, I_m \}$. Proposition \ref{prop:separation from union} implies that $I_1$ and $\bigcup_{i=2}^m I_i $ are separated. Since $S \subseteq I_1 \cup \bigcup_{i=2}^m I_i$, Proposition \ref{prop:connected} then implies that $S \subseteq I_1$ or $S \subseteq \bigcup_{i=2}^m I_i$, but not both. If the former, the result is proved. If the latter, then the same reasoning applied recursively yields the result.
\qed \end{proof}
The next two Lemmas establish that if $\H$ and $\J$ are interval queue representations of the truth sets of formulas $\varphi_1$ and $\varphi_2$ respectively, then $\union \H \boxdot_I \J$ is the truth set of $\varphi_1 \until_I \varphi_2$. 
\begin{lemma} \label{lem:until supercover}
	Let $\H,\J \in \QQ$ be interval queues, $\G$ a finite set of atomic propositions, $I \in \IQ$ nondegenerate, $\varphi_1,\varphi_2 \in \MITL(\G)$, and $z:[0,\infty) \to 2^\G$ a signal of finite variability. If $T_z(\varphi_1) \subseteq \union \H$ and $T_z(\varphi_2) \subseteq \union \J$, then
	$ T_z(\varphi_1 \until_I \varphi_2) \subseteq \union (\H \boxdot_I \J).$
\end{lemma}
\begin{proof}
	Suppose $t \in T_z(\varphi_1 \until_I \varphi_2)$. Then by \eqref{eq:truthSetUntil},
	$ \exists t_2 \in I,\ t + t_2 \in T_z(\varphi_2) \text{ and } (t,t+t_2) \subseteq T_z(\varphi_1). $
	This in turn implies $ \exists t_2 \in I,\ \exists J \in \J,\  t + t_2 \in J \text{ and } (t,t+t_2) \subseteq \union \H.$
	Applying Lemma \ref{lem:subset of IQ} then gives:
	$ \exists t_2 \in I,\ \exists J \in \J,\ \exists H \in \H,\ t + t_2 \in J \text{ and } (t,t+t_2) \subseteq H.$ As such, $[t,t+t_2] \subseteq \Cl(H)$ by Proposition \ref{prop:interiorSub}.
	In particular, $t \in \Cl(H)$, and since $t+t_2 \in J$, we see that $\tau:=t+t_2 \in \Cl(H) \cap J$. Thus $t = \tau - t_2$, with $\tau \in \Cl(H) \cap J$ and $t_2 \in I$, which implies $t \in (\Cl(H) \cap J) \ominus I$, and therefore
	$t \in \Big( (\Cl(H) \cap J) \ominus I \Big) \cap \Cl(H)$. Equation \eqref{eq:grandUnion} then implies $ t \subseteq \union (\H \boxdot_I \J) $.
\qed \end{proof}
\begin{lemma} \label{lem:until subcover}
		Let $\H,\J \in \QQ$ be interval queues, $\G$ a finite set of atomic propositions, $I \in \IQ$ nondegenerate, $\varphi_1,\varphi_2 \in \MITL(\G)$, and $z:[0,\infty) \to 2^\G$ a signal of finite variability. If $T_z(\varphi_1) \supseteq \union \H$ and $T_z(\varphi_2) \supseteq \union \J$, then
	$ T_z(\varphi_1 \until_I \varphi_2) \supseteq \union (\H \boxdot_I \J).$
\end{lemma}
\begin{proof}
	Suppose $t \in \union (\H \boxdot_I \J)$. Then by \eqref{eq:grandUnion},
	$ \exists H \in \H,\ \exists J \in \J,\ t \in \Big( (\Cl(H) \cap J) \ominus I \Big) \cap \Cl(H).$
	It follows that $t \in \Cl(H)$, and
	\begin{equation} \exists \tau \in \Cl(H) \cap J,\ \exists t_2 \in I,\ t = \tau - t_2. \label{eq:tauInJ} \end{equation}
	Since $t \in \Cl(H)$ and $\tau = t + t_2 \in \Cl(H)$ for the interval $H$, it follows that $[t,t+t_2] \subseteq \Cl(H)$. So, by Proposition \ref{prop:interiorSub}, $(t,t+t_2) = \Int([t,t+t_2]) \subseteq \Int( \Cl(H) ), $
	and by Proposition \ref{prop:interiorOfClosure}, $\Int( \Cl(H) ) \subseteq H \subseteq T_z(\varphi_1).$
	Thus, $(t,t+t_2) \subseteq T_z(\varphi_1)$, and $\tau = t + t_2 \in J \subseteq T_z(\varphi_2)$ by \eqref{eq:tauInJ}, where $t_2 \in I$. Finally, $t \in T_z(\varphi \until_I \varphi_2)$ then follows from \eqref{eq:truthSetUntil}.
\qed \end{proof}
\begin{corollary} \label{cor:generalUntil} Under the hypotheses of Lemma \ref{lem:until subcover}, $\H \boxdot_I \J \in \QQ$. Moreover, if $T_z(\varphi_1) = \union \H$ and $T_z(\varphi_2) = \union \J$, then
	$ T_z(\varphi_1 \until_I \varphi_2) = \union (\H \boxdot_I \J).$ Finally, $\H,\J \subseteq \IQ \implies \H \boxdot_I \J \subseteq \IQ$. 
	\end{corollary}
\begin{proof}
	This follows directly from Lemmas \ref{lem:until supercover} and \ref{lem:until subcover}, and the discussion immediately preceding Lemma \ref{lem:subset of IQ}. \qed 
	\end{proof}
\begin{remark}
	Corollary \ref{cor:generalUntil} extends the `General Until' claim in \cite[Section 3]{malerMonitoringTemporalProperties2004} in three directions. First, $I$ needs neither be closed nor bounded. Second, the truth sets of the arguments need not be left-closed right-open. Finally, Corollary \ref{cor:generalUntil} holds for a strict non-matching until, which is more expressive~\cite{furiaExpressivenessMTLVariants2007} than the non-strict matching until of \cite{malerMonitoringTemporalProperties2004}. 
 \end{remark}

\subsection{Constructing under- and over-approximations}
The maps $\P^-$ and $\P^+$ in Problem \ref{prob:relaxed} provide, respectively, under-approximations and over-approximations to the truth sets of the atomic propositions. It is now shown how to combine them, using the operations in Definition \ref{def:interval queue operations}, to obtain under- and over-approximations to the truth sets of any MITL formula, thereby solving Problem \ref{prob:relaxed}. Theorem \ref{thm:consistent interval collections} then states this result formally. 
\begin{definition} \label{def:subcovers and supercovers} Given the two maps $\P^-,\P^+:\G \to \QQ$ in Problem \ref{prob:relaxed}, define $\Q^-,\Q^+: \MITL(\G) \to \QQ$ inductively as follows: for any $\varphi_1,\varphi_2 \in \MITL(\G)$,
	\begin{align}
		&\Q^+(\top) = \{ [0,\infty) \}, &&\Q^-(\top) = \{ [0,\infty) \}, \label{eq:top and bot} \\
		&\Q^+(\g) = \P^+(\g), &&\Q^-(\g) = \P^-(\g), \label{eq:atoms} \\
		&\Q^+(\lnot \varphi_1)  = {\sim} \Q^-(\varphi_1), &&\Q^-(\lnot \varphi_1) = {\sim} \Q^+(\varphi_1), \label{eq:complements} \\
		&\Q^+(\varphi_1 \land \varphi_2) = \Q^+(\varphi_1) \sqcap \Q^+(\varphi_2) , && \Q^-(\varphi_1 \land \varphi_2) = \Q^-(\varphi_1) \sqcap \Q^-(\varphi_2); \label{eq:ands} \end{align}
	and finally for any non-degenerate $I \in \IQ$,
	\begin{align}	& \Q^+(\varphi_1 \until_I \varphi_2) =  \Q^+(\varphi_1) \boxdot_I \Q^+(\varphi_2), && \Q^-(\varphi_1 \until_I \varphi_2) =  \Q^-(\varphi_1) \boxdot_I \Q^-(\varphi_2). \label{eq:untils}
	\end{align}
\end{definition}
\begin{theorem} \label{thm:consistent interval collections}
	Under the hypotheses of Problem \ref{prob:relaxed}, the interval queues $\Q^+(\varphi) \in \QQ$ and $\Q^-(\varphi) \in \QQ$ constructed according to Definition \ref{def:subcovers and supercovers} satisfy
	$$ \union \Q^-(\varphi) \subseteq T_z(\varphi) \subseteq \union \Q^+(\varphi),$$
	for any $\varphi \in \MITL(\G)$. Moreover, if both  $\P^-(\g)\subseteq \IQ$ and $\P^+(\g)  \subseteq \IQ$ for every $\g \in \G$, 
	then for any $\varphi \in \MITL(\G)$, both $\Q^-(\varphi) \subseteq \IQ$ and $\Q^+(\varphi)  \subseteq \IQ$. That is,
	\begin{equation} \bigcup_{\g \in \G}\left(  \P^-(\g) \cup  \P^+(\g) \right) \subseteq \IQ \implies \bigcup_{\varphi \in \MITL(\G)} \left( \Q^-(\varphi) \cup  \Q^+(\varphi) \right) \subseteq \IQ. \label{eq:ER} \end{equation}
\end{theorem}
The proof of Theorem \ref{thm:consistent interval collections} is a straightforward application of all the preceding results, so it is deferred to Appendix \ref{sec:proof}. 
\begin{corollary} \label{cor:soundness}
	For any $\varphi \in \MITL(\G)$, \begin{align*}
		t \in \union \Q^-(\varphi) &\implies (z,t) \models \varphi, 
		\\ t \notin \union \Q^+(\varphi) &\implies (z,t) \not\models \varphi.\end{align*}
\end{corollary}
Thus, in order to test whether an MITL formula is satisfied (or not satisfied) at a given time, we need only look at whether the generated under- and over-approximations contain an interval containing that time.
\begin{remark} The verification result is inconclusive for any $ t \in \union \Q^+(\varphi) \setminus \union \Q^-(\varphi)$. \label{rem:gap} 
\end{remark}
Our procedure also solves Problem \ref{prob:exact} when the under- and over-approximations to the truth sets of the atomic propositions coincide.
\begin{corollary}\label{cor:exact IQs assuming exact props}
If $\P^- = \P^+$, then $\forall \varphi \in \MITL(\G),\ \union \Q^-(\varphi) = T_z(\varphi) = \union \Q^+(\varphi).$
That is, $\P^- = \P^+ \implies \Q^- = \Q^+$.
\end{corollary}
\subsection{Conservativeness}
Corollary \ref{cor:soundness} establishes that our verification procedure is sound. However, due to the uncertainty in initial atomic proposition timing accommodated by Problem \ref{prob:relaxed}, it is necessarily conservative. For a given instance of the problem, the degree of conservativeness relates to the gap between under-approximation $\Q^-(\varphi)$ and over-approximation $\Q^+(\varphi)$, as noted in Remark \ref{rem:gap}. For example, one could use $\Delta := \lambda \left( \union \Q^+(\varphi) \setminus \union \Q^-(\varphi) \right)$, where $\lambda$ is the Lebesgue measure, to quantify conservativeness. The next section illustrates the growth of the gap through the parse tree of a formula by plotting $\Delta$ for each of its sub-formulas. 
A more comprehensive analysis of the gap, and its dependence on the complexity of the formula, is currently in progress. Also related to conservativeness is the notion of time robustness~\cite{donzeRobustSatisfactionTemporal2010}, which quantifies how robust the satisfaction of a formula is to shifts of the signal in time. The precise relationship between time robustness and the gap $\union \Q^+(\varphi) \setminus \union \Q^-(\varphi)$ is a topic of ongoing investigation. 
\section{Numerical Example} \label{sec:numerical}
\begin{figure}[p]
	\includegraphics[width=\textwidth]{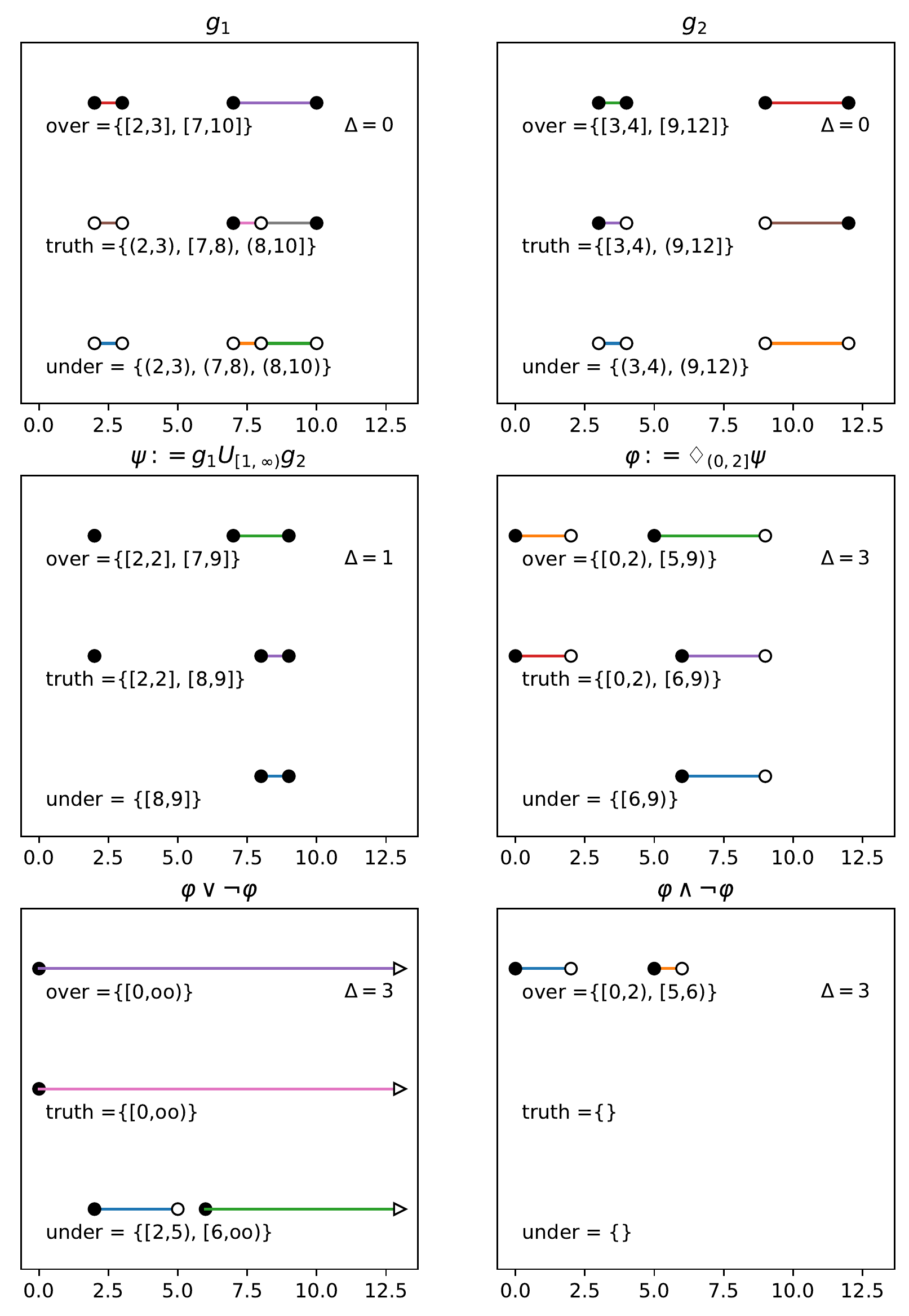}
	\caption{Output of verification algorithm. The interval queues are written explicitly beneath each line. The measure of the gap between the over- and under-approximations is given by $\Delta$. }
	\label{fig:example}
\end{figure}

Here we consider a particular instance of Problem \ref{prob:relaxed} involving two atomic propositions: $\G:=\{\g_1,\g_2\}$. The corresponding truth sets are plotted in the first row of Figure \ref{fig:example}, together with their under- and over-approximations. The under-approximations are  the interiors of the truth sets, and the over-approximations their closures, so initially $\Delta=0$ for both $\g_1$ and $\g_2$ because their under- and over-approximations differ by only a finite number of points. In the next row of Figure \ref{fig:example}, we see the effect of two successive until operators is to increase $\Delta$ by 1 and then again by 2. Note that $\eventually_I \psi := \top \until_I \psi$ is the \emph{eventually} operator. In the third row, we see that this gap remains at $3$, even for the validity $\varphi \lor \lnot \varphi$, and the contradiction $\varphi \land \lnot \varphi$. 
\begin{remark} \label{rem:structure}
	 For some formulas, the verification result can be known for all times simply by virtue of the structure of the formula. However, our procedure may still return an inconclusive result over a subset of times, due to the uncertainty at lower levels of the parse tree. For this reason, $\top$ is chosen as primitive in Definition \ref{def:MITL}, rather than defining $\top:=\g \lor \lnot \g$ in terms of some $\g \in \G$. Then, for contradictions, simply put $\bot:= \lnot \top$. 
 \end{remark}
\section{Conclusion}
Problem \ref{prob:exact} is to verify a continuous-time signal against an MITL formula, given knowledge of the time intervals over which the individual atomic propositions are true. Problem \ref{prob:relaxed} poses a more challenging version of this, in which only over- and under-approximations for those intervals are known. This paper presents a solution to Problem \ref{prob:relaxed}, which also solves Problem \ref{prob:exact} when the intervals are known exactly. 

To formulate and solve Problem \ref{prob:relaxed}, we introduce the notion of an interval queue, which is a finite set of non-empty separated intervals. Separation is the crucial topological property that our procedure relies on. Section \ref{sec:basics} presents algorithms for constructing and complementing interval queues, which become core subroutines within our solution. In Section \ref{sec:solution}, we construct interval queue operators corresponding to each of the primitive logical operators: conjunction, complementation and strict non-matching until. We then show how to combine these operators to obtain interval queue over- and under-approximations to the truth sets of arbitrary MITL formulas. A numerical example in Section \ref{sec:numerical} illustrates this procedure. 

The gap between the over- and under-approximations indicates the conservativeness of the approach. An analysis of the evolution of this gap through the parse tree of the formula is the topic of ongoing work; specifically, of the degree to which reducing the initial gap $\union \P^+(\g) \setminus \union \P^-(\g)$ for every atomic proposition $\g \in \G$, consequently reduces the gap $\union \Q^+(\varphi) \setminus \union \Q^-(\varphi)$ for any formula $\varphi$. This would help assess the benefits of spending more on computation or measurement,  in order to improve the initial timing bounds for the atomic propositions. 
It is also of interest to address Remark \ref{rem:structure}, by considering how information about the structure of the formula can be exploited in order to reduce conservativeness. 
Finally, the relationships between the gap $\union \Q^+(\varphi) \setminus \union \Q^-(\varphi)$ and the notions of temporal robustness in \cite{donzeRobustSatisfactionTemporal2010,lindemannTemporalRobustnessStochastic2022,rodionovaTemporalRobustnessTemporal2022} merit investigation. Recent works~\cite{rodionovaTimeRobustControlSTL2021,linOptimizationbasedMotionPlanning2020} consider the synthesis of controllers to maximise temporal robustness, so such problems may benefit from our interval queue-based analysis.  

%
%
%
\newpage
\bibliographystyle{splncs04}
\bibliography{../References/FormalMethodsInControl}

\appendix
	\section{Proof of Theorem \ref{thm:consistent interval collections}}
	\label{sec:proof} 
We first show the result holds for $\varphi \in \{\top\} \cup \G$.
\begin{itemize}
	\item By \eqref{eq:top and bot}, $\Q^-(\top) = \Q^+(\top) = \{ [0,\infty)\} \in \QQ$, and therefore $\union \Q^-(\top) = [0,\infty) = T_z(\top) = \union \Q^+(\top)$.
	\item By \eqref{eq:relaxedPs}, $\Q^-(\g),\Q^+(\g) \in \QQ$, and $\union \Q^-(\g) \subseteq T_z(\g) \subseteq \union \Q^+(\g)$ for all $\g \in \G$ by \eqref{eq:atoms}. 
\end{itemize}
For the remaining steps, suppose that that $\Q^-(\varphi_1), \Q^+(\varphi_1), \Q^-(\varphi_2), \Q^+(\varphi_2) \in \QQ$ are interval queues such that
\begin{align} 
	\union \Q^-(\varphi_1) &\subseteq T_z(\varphi_1) \subseteq \union \Q^+(\varphi_1), \label{eq:phi1} \\ 
	\union \Q^-(\varphi_2) &\subseteq T_z(\varphi_2) \subseteq \union \Q^+(\varphi_2). \label{eq:phi2} 
\end{align}
\begin{itemize}
	\item Lemma \ref{lem:complement IQ} implies $\Q^-(\lnot \varphi_1) = \mathtt{ComplementIQ}( \Q^+(\varphi_1) ) \in \QQ$. If $t \in \union \Q^-(\lnot\varphi_1)$, then $t \notin \union \Q^+(\varphi_1)$ by Lemma \ref{lem:complement IQ}, and \eqref{eq:phi1} then implies $t \notin T_z(\varphi_1)$. It follows that $(z,t) \not \models \varphi_1$, by which $(z,t) \models \lnot \varphi_1$, and therefore $t \in T_z(\lnot \varphi_1)$. Thus, $\union \Q^-(\lnot\varphi_1) \subseteq T_z(\lnot \varphi_1)$. 
	\item  Lemma \ref{lem:complement IQ} implies $\Q^+(\lnot \varphi_1) = \mathtt{ComplementIQ}( \Q^-(\varphi_1) ) \in \QQ$. If $t \in T_z(\lnot \varphi_1)$, then $t \notin T_z(\varphi_1)$, and \eqref{eq:phi1} then implies $t \notin \union \Q^-(\varphi_1)$. Therefore $t \in [0,\infty) \setminus \union \Q^-(\varphi_1) = \union \Q^+(\lnot \varphi_1)$ by Lemma \ref{lem:complement IQ}. Thus, $T_z(\lnot \varphi_1) \subseteq \union \Q^+(\lnot \varphi_1)$.
	\item Lemma \ref{lem:intersection IQ} implies $\Q^-(\varphi_1 \land \varphi_2) = \Q^-(\varphi_1)\sqcap \Q^-(\varphi_2) \in \QQ$. Furthermore, \eqref{eq:phi1} - \eqref{eq:phi2} imply
	$\union \Q^-(\varphi_1 \land \varphi_2) = \union (\Q^-(\varphi_1) \sqcap \Q^-(\varphi_2)) =  \union \Q^-(\varphi_1) \cap \union \Q^-(\varphi_2) \subseteq T_z(\varphi_1) \cap T_z(\varphi_2) = T_z(\varphi_1 \land \varphi_2)$. 
	\item Similarly, $\Q^+(\varphi_1 \land \varphi_2) \in \QQ$, and $T_z(\varphi_1 \land \varphi_2) = T_z(\varphi_1) \cap T_z(\varphi_2) \subseteq \union \Q^+(\varphi_1) \cap \union \Q^+(\varphi_2) = \union (\Q^+(\varphi_1) \sqcap \Q^+(\varphi_2)) = \union \Q^+(\varphi_1 \land \varphi_2)$.
	\item Corollary \ref{cor:generalUntil} implies $\Q^-(\varphi_1 \until_I \varphi_2) = \Q^-(\varphi_1) \boxdot_I \Q^-(\varphi_2) \in \QQ$ for any nondegenerate $I \in \IQ$, and $\union \Q^-(\varphi_1 \until_I \varphi_2) = \union \left( \Q^-(\varphi_1) \boxdot_I \Q^-(\varphi_2) \right)\subseteq T_z(\varphi_1 \until_I \varphi_2)$ by Lemma \ref{lem:until subcover}. 
	\item Corollary \ref{cor:generalUntil} implies $\Q^+(\varphi_1 \until_I \varphi_2) = \Q^+(\varphi_1) \boxdot_I \Q^+(\varphi_2) \in \QQ$ for any nondegenerate $I \in \IQ$, and $\union \Q^+(\varphi_1 \until_I \varphi_2) = \union \left( \Q^+(\varphi_1) \boxdot_I \Q^-(\varphi_2) \right)\supseteq T_z(\varphi_1 \until_I \varphi_2)$ by Lemma \ref{lem:until supercover}. 
\end{itemize}
The result then follows by structural induction on the formula $\varphi$. Endpoint rationality is preserved \eqref{eq:ER} because all the individual operations preserve it. \qed
\end{document}